\documentclass[12pt]{article}
\usepackage{amsmath,amssymb,graphicx,bm}
\usepackage{epsfig}
\usepackage [small] {caption}
\newcommand{\ra}{\rightarrow}
\newcommand{\bq}{\begin{eqnarray}}
\newcommand{\eq}{\end{eqnarray}}
\newcommand{\ov}{\overline}
\newcommand{\beq}{\begin{equation}}
\newcommand{\eeq}{\end{equation}}

\usepackage[english]{babel}
\usepackage{amsfonts}
\usepackage{color}

\begin{document}

\begin{center}
\Large \bf{ Exclusive\,\, $\bm {\gamma^{(*)}\gamma}$\,\,  processes}
\end{center}
\vspace{0.2 cm}

\begin{center}
V.L. Chernyak
\end{center}
\begin{center}
Budker Institute of Nuclear Physics, 630090 Novosibirsk, Russia
\end{center}
\vspace{1mm}
\begin{center}
Talk given at the Beijing Conference\,\, "From  $\phi \,\, to\,\, \psi$\,",\,\, 13 -16 \, October 2009, \, China
\end{center}
\begin{center}
{\bf Abstract}
\end{center}
\begin{center}
A short review of experimental and theoretical results on the large angle cross sections \\
$"\gamma\gamma\to {\rm two\,mesons}"$ and the form factors $\gamma^*\gamma\to P=\{\pi,\,\eta,\, \eta'\}$ is given.
\end{center}

\vspace*{5mm}

\begin{center}
\bf \large 1. \quad Introduction
\end{center}

The general approach to calculations of hard exclusive processes in QCD was developed in \cite{cz1, cz2}. In particular,
the general formula for the {\it leading\,\, power\, term of any hadron form factor} $\gamma^*\to H_1 H_2$
has the form \cite{cz1} \,:
\beq
\langle p_1,\,s_1,\,\lambda_1;\, p_2,\,s_2,\,\lambda_2|J_{\lambda}|0\rangle =  C_{12}
\Bigl (1/\sqrt {q^2}\Bigr )^{|\lambda_1+\lambda_2|+(2n_{min}-3)}\,,
\eeq
where\,: $n_{\min}$ is the minimal number of elementary constituents in a given
hadron,\, $n_{min}=2$  for mesons and $n_{min}=3$  for baryons\,;\,\,
$s_{1,2}$ and $\lambda_{1,2}$ are the hadron spins and helicities,
the current helicity $\lambda=\lambda_1-\lambda_2=0,\,\pm 1$\,;
the coefficient $C_{12}$ is expressed through the integral over the wave functions of both hadrons.
\vspace{1mm}

It is seen that the behavior is {\it  independent of hadron spins, but depends essentially on their helicities},
and the {\it QCD helicity selection rules} are clearly  seen:  the largest form factors occur only
for $\lambda_1=\lambda_2=0$ mesons and $\lambda_1=-\lambda_2=\pm 1/2$ baryons of {\it any spins}.
\vspace{1mm}

The QCD logarithmic loop corrections to (1) were first calculated in \cite{cz2} (see also \cite{cz3}\cite{ER}\cite{BL1}
for more details and \cite{cz-r} for a review).
\vspace{-2mm}
\begin{center}
{\bf 2.} \quad $\Large {\bm{\gamma\gamma\to {\ov M}M} \quad  \bf large\,\, angle\,\, scattering}$
\end{center}

The QCD predictions for the {\it leading terms} of the large angle scattering cross sections
$\gamma\gamma \to {\rm two \,\, mesons}$ were considered in \cite{BL2}\cite{Maurice} (see also \cite{DN}
for the one-loop corrections)

\begin{minipage}{0.8\textwidth}
\vspace{0.3 cm}
\centering
\includegraphics[width=0.45\textwidth]{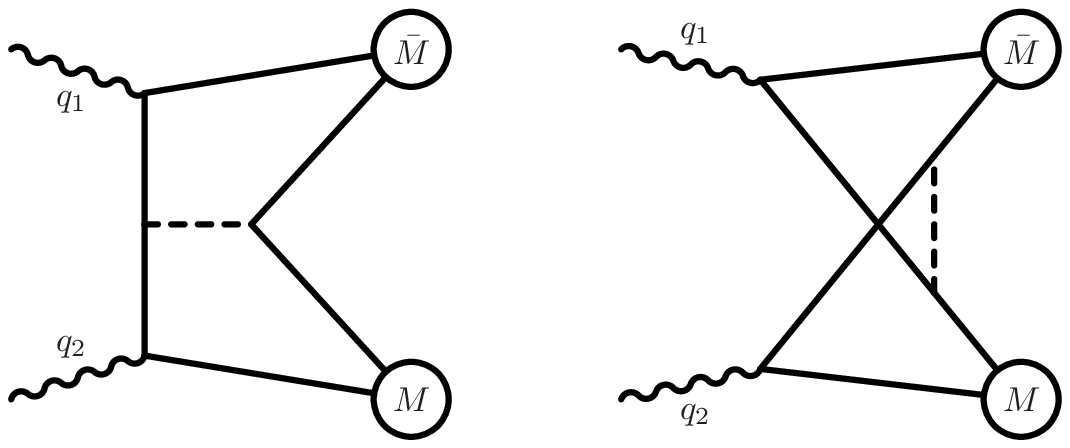}
\end{minipage}
\vspace{0.2 cm}

\begin{scriptsize}
\begin{center}
Fig.1 \,\, Two typical lowest order Feynman diagrams for the leading term hard QCD contributions\\ to
$\gamma\gamma\ra {\ov M}M$\,, the broken line is the hard gluon exchange.
\end{center}
\end{scriptsize}

The expressions for the cross sections look as (the example in (2) is given for $\gamma\gamma\to K^+K^-$)\,:
\beq
\frac{d\sigma(\gamma\gamma\ra M^{\dagger}\,M)}{d\cos\theta}=\frac{1}{32\pi W^2}\,\frac{1}{4}\sum_{
\lambda_1 \lambda_2}\Bigl | A_{\lambda_1\lambda_2}\Bigr |^2\,, \nonumber
\eeq
\vspace{-5mm}
\beq
A^{(lead)}_{\lambda_1\lambda_2}(W,\theta)=\frac{64\pi^2}{9W^2}\,\alpha \,{\ov \alpha}_s \,f_P^2
\int_0^1 dx\, \phi_{P}(x)\int_0^1 dy\, \phi_{P}(y)\,T_{\lambda_1\lambda_2}
(x,\,y,\,\theta)\,,\nonumber
\eeq
\vspace{-3mm}
\beq
T_{++}=T_{--}=(e_u-e_s)^2\,\frac{1}{\sin^2\theta}\,\frac{A}{D}\,,
\eeq
\vspace{-5mm}
\beq
T_{+-}=T_{-+}=\frac{1}{D}\Biggl [ \frac{(e_u-e_s)^2}{\sin^2\theta}(1-A)+e_u e_s \frac
{A C}{A^2-B^2\cos^2\theta}+\frac{(e_u^2-e_s^2)}{2}(x_u-y_s)  \Biggr ],\nonumber
\eeq
\beq
A=(x_sy_u+x_uy_s)\,,\,\quad B=(x_sy_u-x_uy_s)\,,\,\quad C=(x_sx_u+y_sy_u)\,,\quad D=x_u x_sy_u y_s\,,\nonumber
\eeq
where: $x_s+x_u=1\,,\quad e_u=2/3, \quad e_s=e_d=-1/3\,,\,\,  f_P$ are the couplings\,:\,
$f_{\pi}\simeq 132\,{\rm MeV}\,,\,\, f_K\simeq 162\,{\rm MeV}\,,\,\, \phi_P(x)$ is the leading twist pseudoscalar meson wave
function (= distribution amplitude), $"x"$ is the meson momentum fraction carried by quark inside the meson.
\vspace{4mm}

\begin{minipage}{0.75\textwidth}
\centering
\includegraphics[width=0.4\textwidth]{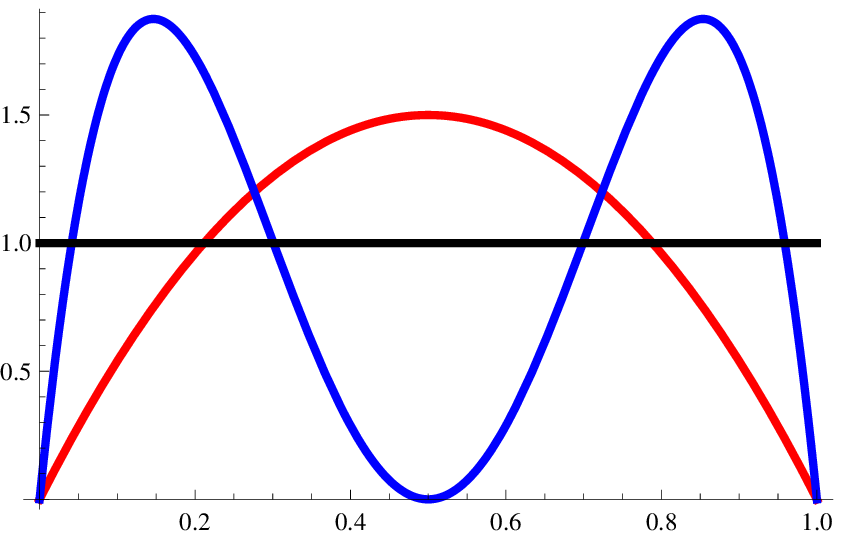}
\end{minipage}
\vspace{2mm}

\begin{scriptsize}
\begin{center}
Fig.2\,\,\,  Three different models for the leading twist  pion wave function $\phi_{\pi}(x)$.\\
Red line -  asymptotic wave function $\phi_{\pi}^{\rm asy}(x)=6x(1-x)$\,.\,\,\\
Blue line -  CZ wave function (at the low scale normalization point $\mu_o\sim 1\,GeV$)\, :
$\phi^{cz}_{\pi}(x,\mu_o)=30x(1-x)(2x-1)^2$\,~ \cite{cz}\,.\\
Black line -  flat wave function $\phi_{\pi}(x,\mu_o\sim 1\,GeV)=1$.
\end{center}
\end{scriptsize}

Cross sections for {\it charged\, mesons}\,: $\gamma\gamma \to \pi^{+}\pi^{-},\,\, K^{+}K^{-}$\, behave as:
\bq
\frac{d\sigma(\gamma\gamma\to \pi^{+}\pi^{-})}{d\cos\theta}\sim \frac{f_{\pi}^4}{W^{6}\sin^{4}\theta}\, ,
\eq
and the angular distribution $\sim 1/\sin^{4}\theta$ is only weakly dependent of the meson wave function form.
But {\it the absolute values of cross sections depend strongly on the form of $\phi_M(x)$} and are much larger for the
wide wave functions.
\vspace{1mm}

For {\it neutral\, mesons}\,:\, $\gamma\gamma\ra \pi^{o}\pi^{o},\,{\ov K}_S K_S,\, \pi^{o}\eta,\, \eta\eta$\,\, the
coefficient of the formally leading term $\sim 1/W^6$ is {\it very small}, so that at present energies $W< 4\,GeV$
such amplitudes are dominated by the first power correction in the amplitude and the energy behavior is much steeper:
\bq
\frac{d\sigma(\gamma\gamma\ra {\ov K}_S K_S)}{d\cos\theta}\sim \frac{f_{K}^4}{W^{10}}\,\chi(\theta)\,,
\eq
while, unlike (3), the angular dependence $\chi(\theta)$ and the overall coefficient in (4) are not predicted
(at present) in a model independent way.
\vspace{2mm}

As the alternative approach to description of $\gamma\gamma\to {\ov M}M$ processes, the "handbag model" was
used in \cite{DKV}. The main dynamical assumption of the "handbag model"\, \cite{DKV} is that at present
energies $W\leq 4\, GeV$ {\it all}\, $\gamma\gamma\to {\ov M}M$  amplitudes are still dominated by soft non-leading terms.

\begin{minipage}{1.0\textwidth}
\vspace{2mm}
\centering
\includegraphics[width=0.4\textwidth]{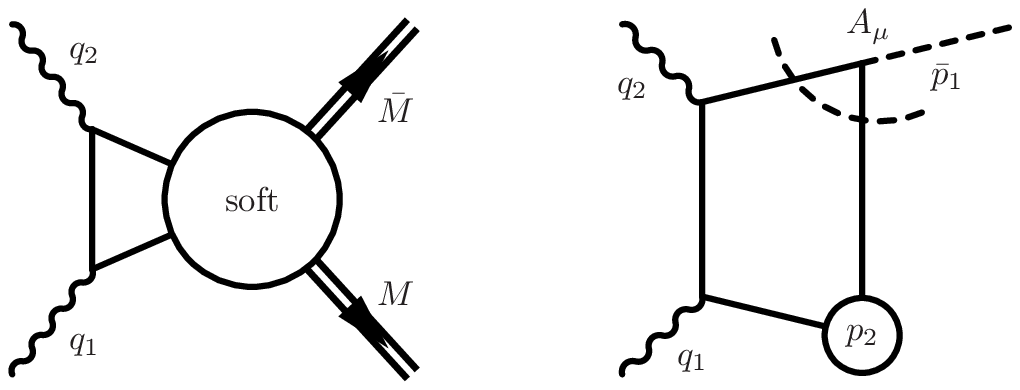}
\end{minipage}

\begin{scriptsize}
\begin{center}
Fig.3a\,\, The overall picture of the "standard handbag" contribution \cite{DKV}.\\
\end{center}
\vspace*{-4mm}
\begin{center}
Fig.3b\, The standard lowest order Feynman diagram for the QCD light cone sum rule \cite{Ch1}.
\end{center}
\end{scriptsize}

For {\it all\, mesons}, both charged and neutral, "the standard handbag" contribution (Fig.3) gives:\,
$d\sigma(\gamma\gamma\to {\ov M} M)/d\cos\theta \sim {\rm const}/W^{10}$\, \cite{Ch1}. This angular behavior
${\it \sim const}$ disagrees with all data $\sim 1/\sin^{4}\theta$, and the energy behavior
disagrees with the data $\sim 1/W^6$ for charged mesons.
\footnote{\,
It was "obtained" in \cite{DKV} that the angular behavior of the standard handbag contribution from Fig.3a is $d\sigma/
d\cos\theta\sim 1/\sin^{4}\theta$. Really, this "result" is completely model dependent. The reason is that a number of
special approximate relations were used in \cite{DKV} at intermediate steps. All these relations are valid, at best,
for the leading term only. But it turned out finally that their would be leading term gives zero contribution to the
amplitude, and the whole answer is due to next corrections, which were not under control in \cite{DKV}. Their result
$\sim 1/\sin^{4}\theta$  is completely due to especially (and arbitrary) chosen form of the next to
leading correction, while ignoring all others next to leading corrections of the same order of smallness. Therefore,
there is no really model independent prediction of the angular dependence in \cite{DKV}. So, it is not surprising that
the explicit calculation in \cite{Ch1} gives different angular dependence.
}
\vspace{1mm}

\begin{minipage}[c]{.5\textwidth}\hspace*{1cm}
\includegraphics
[trim=0mm 0mm 0mm 0mm, width=0.4\textwidth,clip=true]{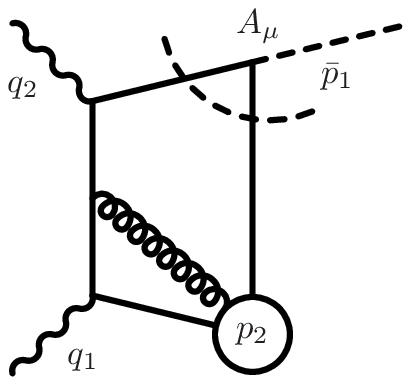}
\end{minipage}~
\begin{minipage}[c]{.5\textwidth}\hspace*{-1cm}
\begin{footnotesize}
\hspace*{-3cm} Fig.4\,\quad  The typical additional Feynman diagram \\
\hspace*{-4.5cm}  for the "extended handbag model"  which includes contributions\\
\hspace*{-4.5cm}  from 3-particle wave functions (the curly line is the near mass\\
\hspace*{-2.5cm} shell non-perturbative gluon).
\end{footnotesize}
\end{minipage}
\vspace{1mm}

I expect that, in distinction with the standard contribution of Fig.3, the additional soft contributions like those in
Fig.4 will give\,:
\beq
\Biggl (\frac{d\sigma(\gamma\gamma\ra {\ov M}M)}{d\cos\theta}\Biggr )_{fig.3}\sim \frac{1}{W^{10}};\,\,\,
\quad \Biggl (\frac{d\sigma(\gamma\gamma\ra {\ov M}M)}{d\cos
\theta}\Biggr )_{fig.4}\sim \frac{1}{W^{10}\sin{^4}\theta}\,\,,
\eeq
in better qualitative agreement with data for {\it neutral\, mesons}. Unfortunately, such contributions are not yet
calculated at present (and it well may be that they are too small in absolute values).

\vspace{2mm}
Now, about a comparison with the data. The Belle results for $\gamma\gamma \ra \pi^+\pi^-$ and $\gamma\gamma\ra K^+K^-$\,\,
\cite{Nakaz} are presented in Fig.5.\\
\vspace{1mm}

\begin{minipage}[c]{.4\textwidth}\vspace*{-4.5mm}\hspace*{-1.5cm}
\includegraphics[trim=0mm 0mm 0mm 0mm, width=0.8\textwidth,clip=true]{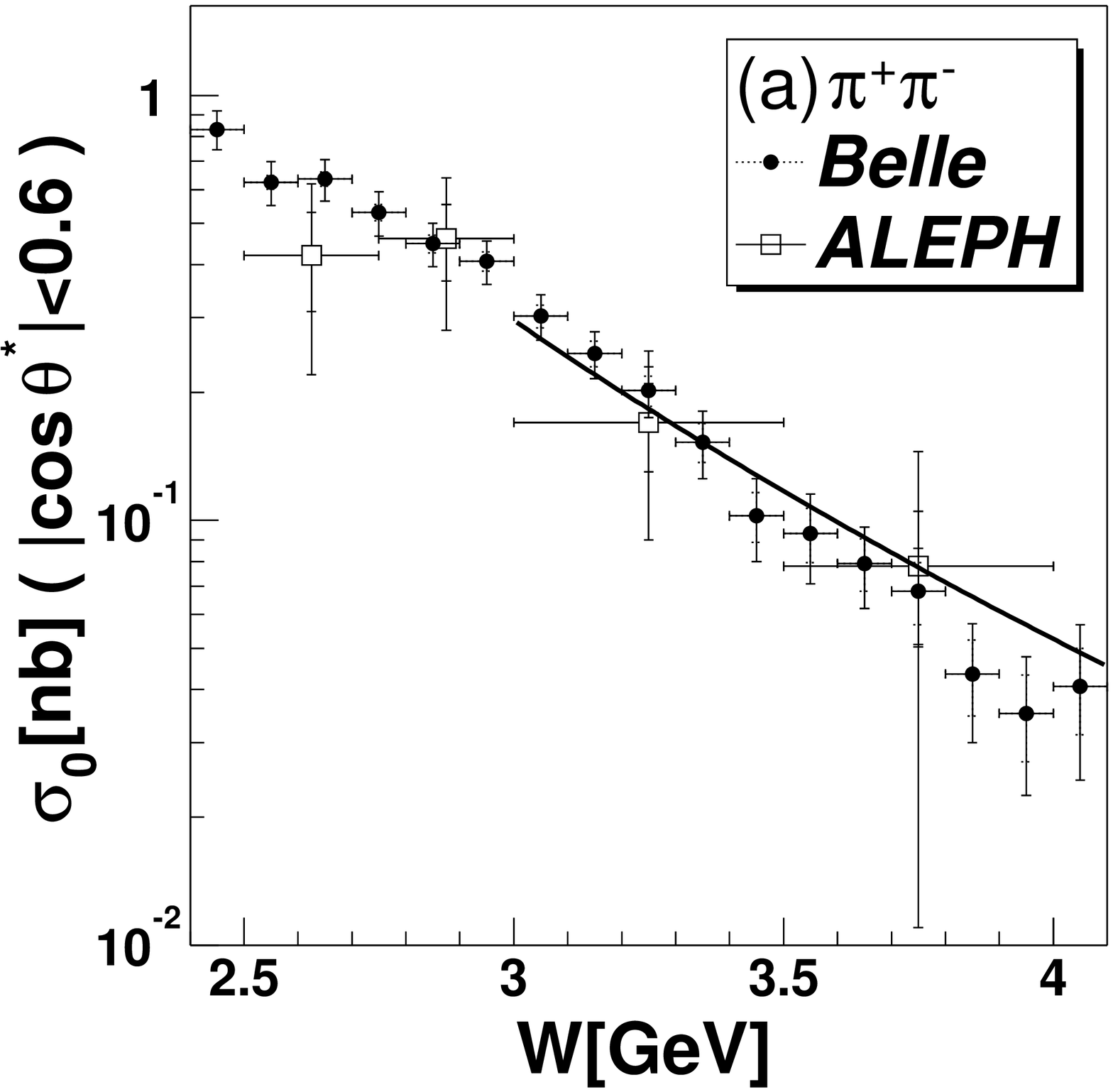}
\end{minipage}~
\begin{minipage}[c]{.4\textwidth}\vspace*{-3mm}\hspace*{-2.9cm}
\includegraphics[trim=0mm 0mm 0mm 0mm, width=0.8\textwidth,clip=true]{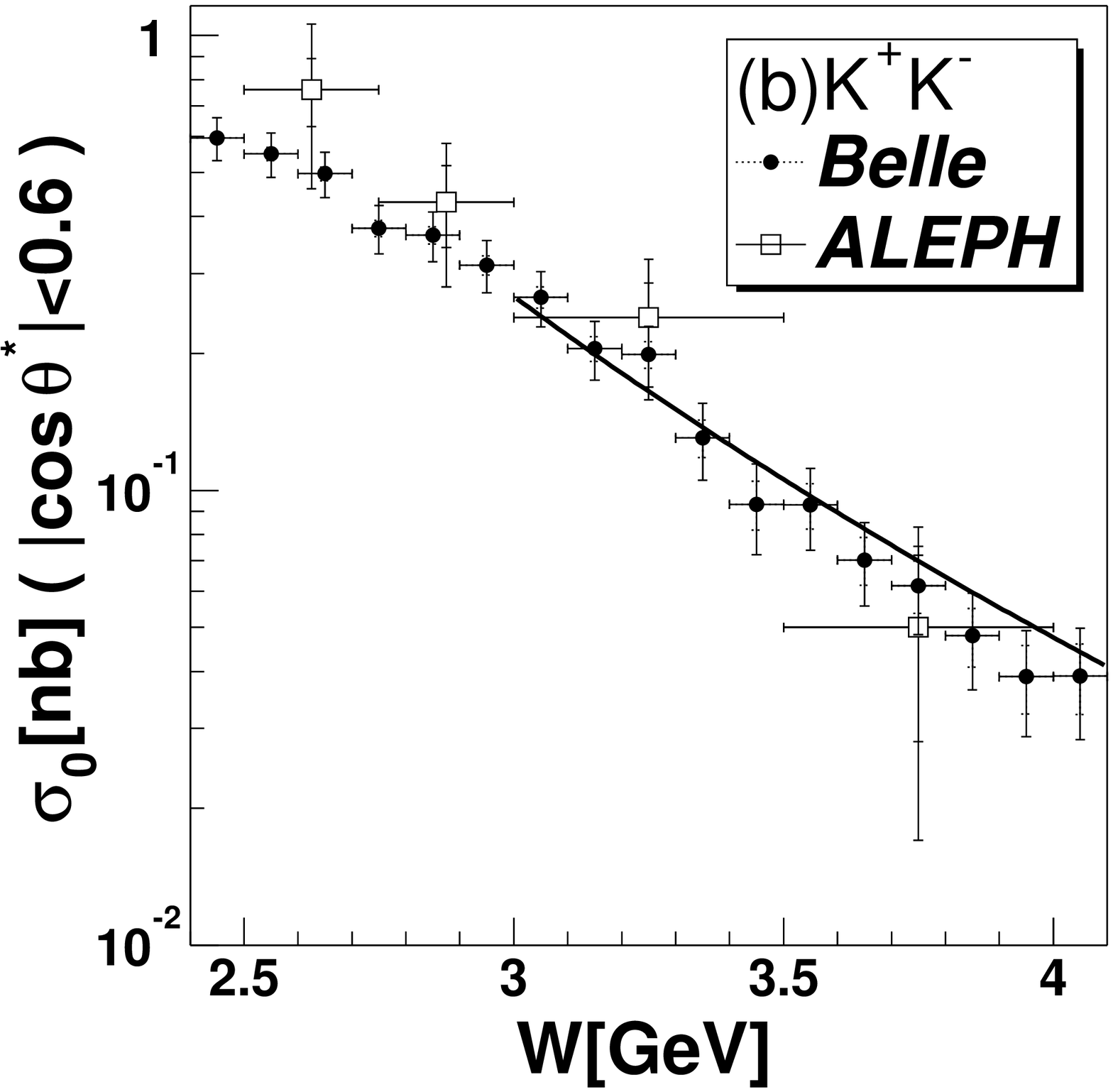}
\end{minipage}
\begin{minipage}[c]{0.4\textwidth}\vspace*{-4mm}\hspace*{-3.9cm}
\includegraphics[width=0.9\textwidth]{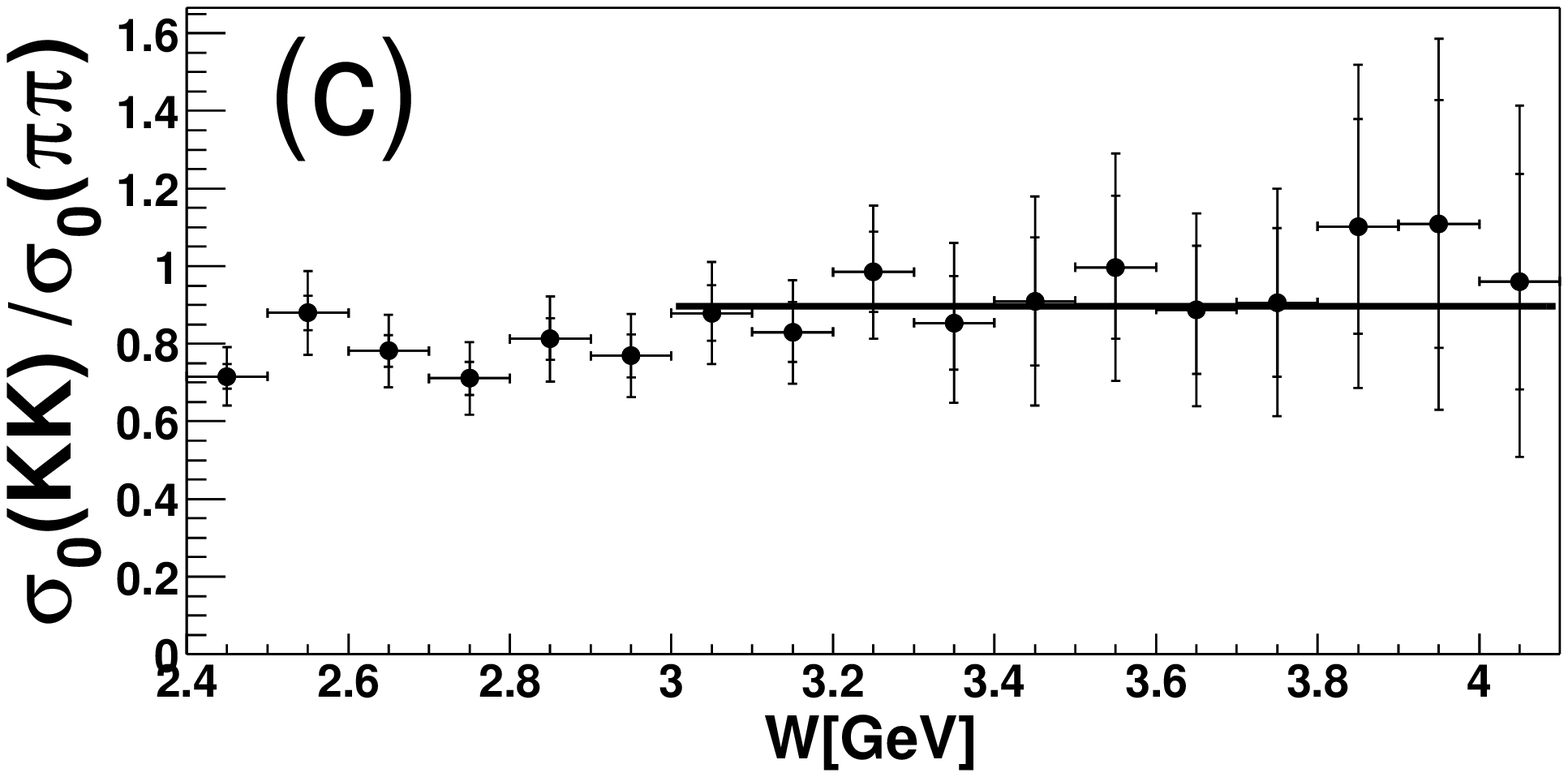}
\end{minipage}~

\begin{scriptsize}
\vspace*{1mm}\hspace*{-4mm}
Fig.5\,\, a,\,b)\, Cross\, sections\,\, \,$\sigma_{o}$\,\,\, integrated over the angular region $|\cos\theta|<0.6$\,,
together with the $\sim (1/W)^6$ dependence line \cite{Nakaz}.\\
\hspace*{-3mm} c)\, the cross section {\it ratio}\, $R_{\rm exp}=\sigma_{o}(K^+K^-)/\sigma_{o}(\pi^+\pi^-)\simeq 0.9$\,
\cite{Nakaz}.\,Compare  $R_{\rm exp}\simeq 0.9$ with the naive prediction\, $R=(f_K/f_{\pi})^4\simeq 2.3$\,.~
\end{scriptsize}
\footnote{\,
The reason for $R_{\rm exp}\ll (f_K/f_{\pi})^4$ is that the leading twist pseudoscalar meson wave function $\phi_P(x)$ becomes
narrower when the lighter $u$ or  $d$ quarks are replaced with the heavier $s$ quarks, and this opposite effect
compensates those from $f_K/f_{\pi}>1$, see \cite{Maurice}.
}
\vspace{2mm}

It is seen that they are compatible with the leading term QCD predictions $\sigma(\pi^+\pi^-)\sim \sigma(K^+K^-)\sim 1/W^6$\,,
and disagree with the handbag model predictions of much steeper behavior $\sim 1/W^{10}$.

\vspace*{1mm}
\begin{minipage}[c]{.5\textwidth}\includegraphics
[trim=0mm 0mm 0mm 5mm, width=0.55\textwidth,clip=true]{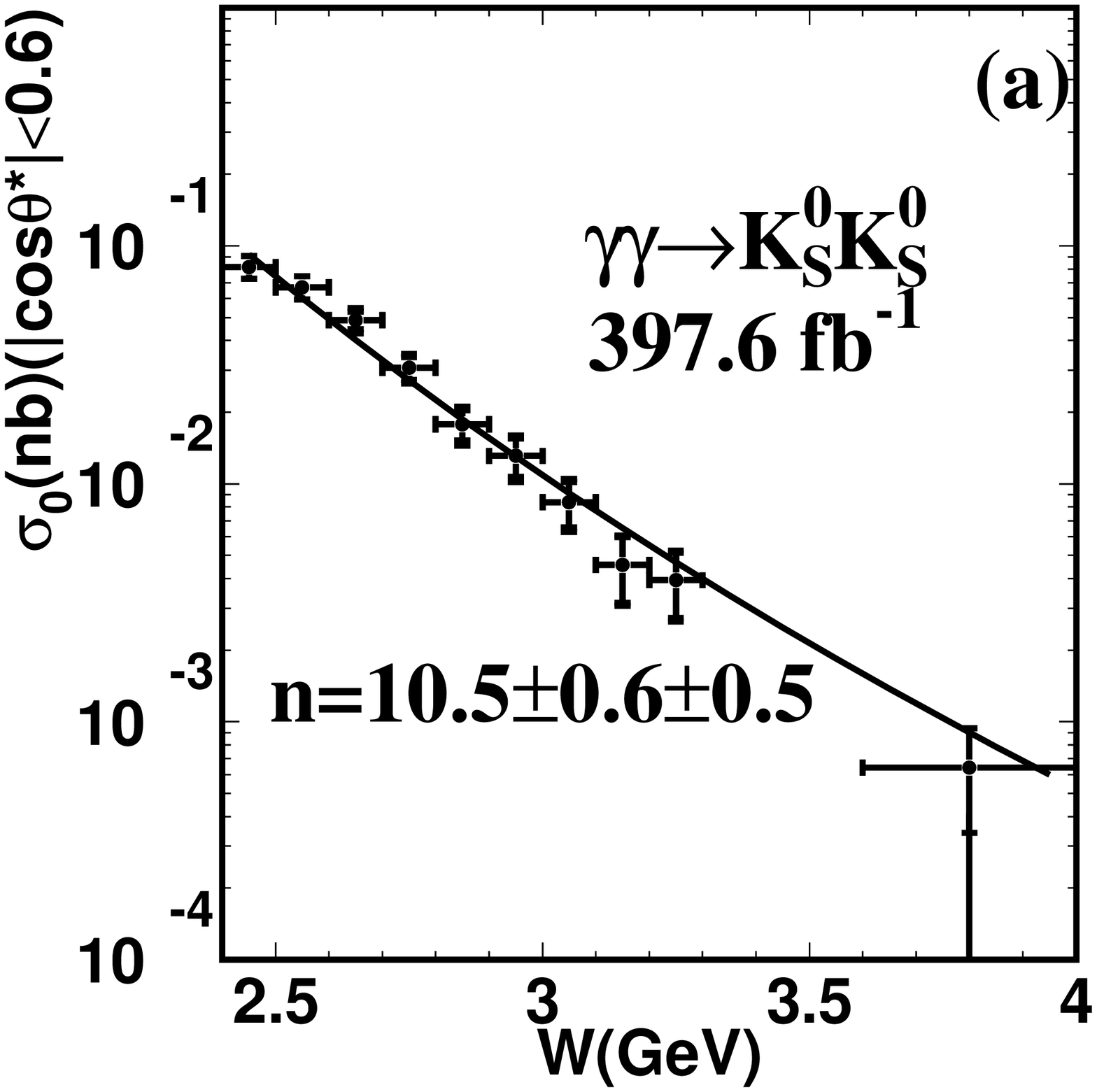}
\end{minipage}~
\begin{minipage}[c]{.5\textwidth}\includegraphics
[trim=0mm 0mm 0mm 5mm, width=0.55\textwidth,clip=true]{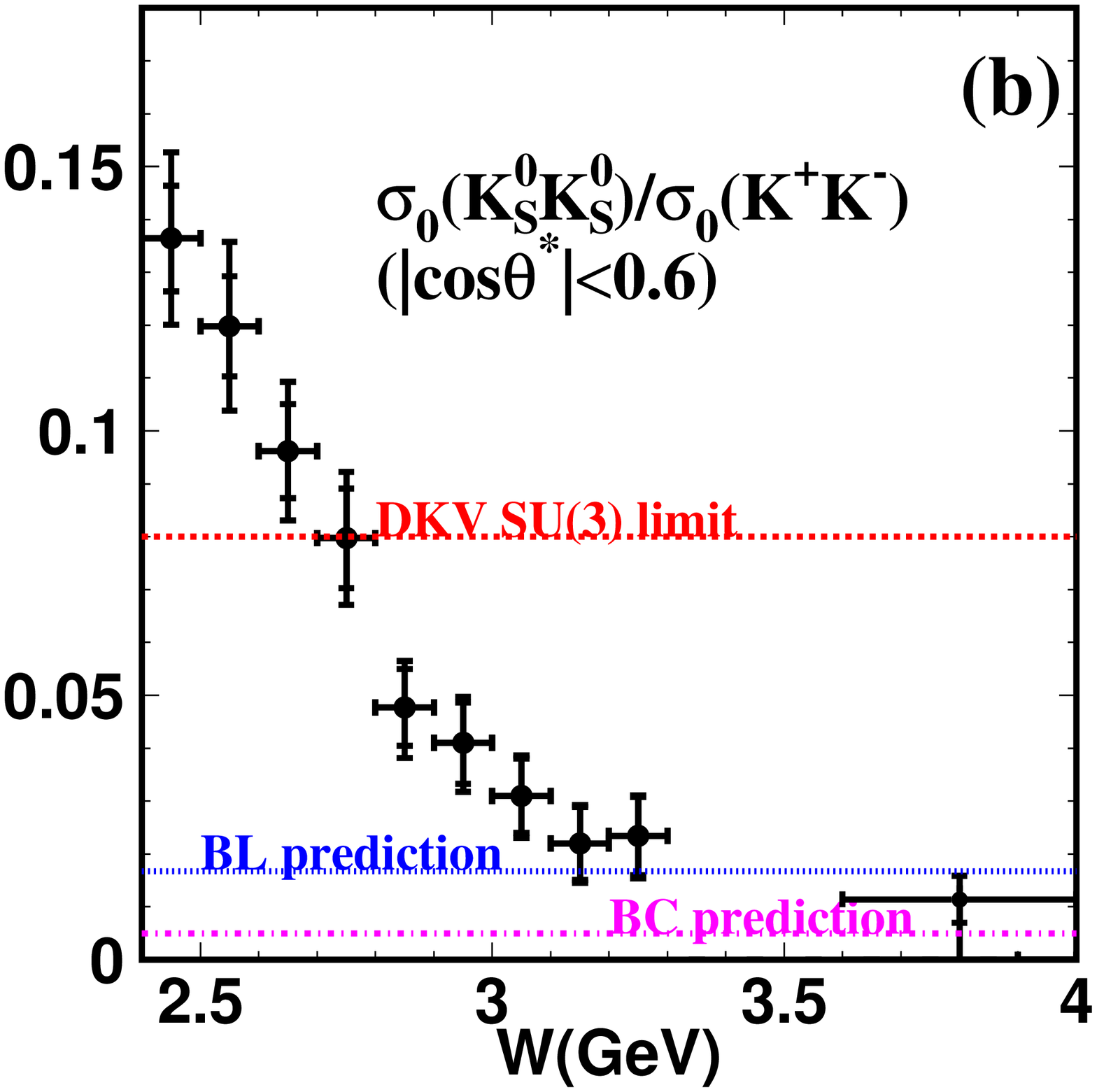}
\end{minipage}
\vspace{0.2 cm}
\begin{scriptsize}\hspace*{3mm}
Fig.6 \,\quad (a)\,\,  The\, total\, cross\, section\,\, $\sigma_o(\gamma\gamma\ra K_SK_S)$ integrated over the angular region
$|\cos\theta|<0.6$\, \cite{Chen}.\\ Here $\bf n$ is the $W$-dependence $\bm{ \sigma_{o}(W)
\sim 1/W^{n}}$; \,\,\,(b)\,\, The  ratio\, $\sigma_0(K_S K_S)/\sigma_0(K^+K^-)$ versus $W$.

The dotted line DKV = Diehl-Kroll-Vogt is the handbag model prediction in the $SU(3)$
-flavor symmetry limit~\cite{DKV}; the dashed BL line is the Brodsky-Lepage \cite{BL2} prediction for the
kaon wave function close to $\phi_{\rm asy}(x)$,\,\, the dashed-dotted BC line is the
Benayoun-Chernyak \cite{Maurice} prediction for the kaon wave function like $\phi_{\rm cz}(x)$ (both are the leading term
QCD predictions (2) for  large energies $W$).
\end{scriptsize}
\vspace{2mm}

The angular behavior measured by Belle \cite{Nakaz} for $\pi^+\pi^-$ and $K^+K^-$ is $\sim 1/\sin^4\theta$\,, also in
agreement with QCD and in disagreement with the standard handbag model \cite{Ch1}. As for
the absolute values of cross sections, the values predicted from (2) are {\it much smaller} than data for the pion
(kaon) wave functions close to $\phi_{\rm asy}(x)$, while predictions from (2) for the wide $\pi$ and $K$ wave functions like
$\phi_{\rm cz}(x)$ are in {\it a reasonable agreement} with data (see \cite{Ch1} for more details).
\vspace{1mm}

Now, let us compare with the Belle results for the {\it neutral mesons}. The results for the cross section $\gamma\gamma\to
K_S K_S$ are published in \cite{Chen}, see Fig.6.

It is seen from Fig.6 that in the energy range $2.5<W<4\,GeV$ the energy behavior $\sim 1/W^{10}$ in this neutral channel
is much steeper in comparison with $\sim 1/W^{6}$ in the charged channel. This agrees with qualitative expectations from QCD
that because the coefficient of the formally leading at sufficiently large $W$ is very small, the first non-leading term
dominates the $K_S K_S$-amplitude at present energies $W<4\, GeV$. Let us recall that the handbag model predicts the dominance of
non-leading terms (and so the energy behavior $\sim 1/W^{10}$) {\it for all mesons}, both charged and neutral.

As for the angular distribution, the data are sufficiently well described by $\sim 1/\sin^{4}\theta$\, \cite{Chen}. Let us
recall once more that "the standard handbag model" (Fig.3) predicts the flat angular distribution $\sim \rm const$
\cite{Ch1} also for all mesons,
but the qualitative expectation is that "the extended handbag model" (Fig.4) will give $\sim 1/\sin^{4}\theta$, see (5).

Finally, about the ratio $R=\sigma_0(K_S K_S)/\sigma_0(K^+K^-)$, see Fig.6b. In the $SU(3)$ flavor symmetry limit the
standard (and extended) handbag model predicts $R=0.08$ \,\cite{DKV}. It is seen from Fig.6b that this ratio decreases
rapidly with energy and becomes smaller than $\sim 0.08$ at $W>2.7\,GeV$, in disagreement with the handbag model. This is
because the energy dependence of $\sigma_0(K^+K^-)\sim 1/W^6$ disagrees with the handbag model prediction $\sim 1/W^{10}$.

The QCD prediction is that at sufficiently large $W$, when the parametrically leading but having a small coefficient term will
become dominant in the $K_S K_S$ amplitude, this ratio will become constant (see BL and BC lines in Fig.6b). It is seen from
Fig.6b that the ratio $R$ is already close to the leading term QCD predictions for $K_SK_S$ at $W\simeq 4\, GeV$.

The qualitative situation with other neutral modes, $\gamma\gamma\to \pi^o\pi^o,\,\pi^o\eta,\, \eta\eta,\, \eta'\eta'$, etc.,
is similar to those of $\gamma\gamma\to K_S K_S$. Recently, there appeared new data from the Belle Collaboration on cross
sections $\gamma\gamma\to \pi^o\pi^o$ and $\gamma\gamma\to \pi^o\eta$ \cite{Ue-pi}\cite{Ue-eta}, see Figs.7,8 and Table 1. The
QCD predictions for this range of energies are: $\sigma(\pi^+\pi^-)\sim 1/W^6\,,\,\,\sigma(\pi^o\pi^o)\sim\sigma(\pi^o\eta)
\sim 1/W^{10},\,\, R=\sigma(\pi^o\pi^o)/\sigma(\pi^+\pi^-)\sim 1/W^4.$ The handbag model prediction is\,: $R=\sigma(\pi^o\pi^o)
/\sigma(\pi^+\pi^-)=1/2\,.$ As for the cross section $\sigma(\gamma\gamma\to \pi^o\eta)$, it behaves "normally",\, $\sim
1/W^{10}$, similarly to $\sigma(\gamma\gamma\to K_S K_S)$, see Fig.8a. But as for $\sigma(\gamma\gamma\to \pi^o\pi^o)$, it
behaves "abnormally"\,,  see Fig.7. This last behavior agrees neither with QCD, nor with the handbag model.

\begin{center}
{\bf 3.} \quad {\bf Conclusions\, on \,the\, large\, angle\, cross\, sections} $\bm{\gamma\gamma\to {\ov M}M}$
\end{center}

1)\,\, The leading term QCD predictions $d\sigma/d\cos\theta\sim 1/(W^6\sin^{4}\theta)$ for charged mesons
$\pi^{+}\pi^{-},\,K^{+}K^{-}$ agree sufficiently well with data both in energy and angular dependence at energies
$W \gtrsim\, 3\, GeV$. The absolute\, values of cross sections are in a reasonable agreement with data {\it only for
the wide $\pi$ (K) wave functions}, like $\phi_{\pi,K}^{cz}(x)$ (see \cite{Maurice,Ch1} for more details). The asymptotic wave
functions $\phi_{\pi,K}(x)\simeq\phi^{asy}(x)$ predict {\it much smaller} cross sections. The handbag
model predictions for charged mesons disagree with data in energy dependence.

2)\,\, For\, neutral\, mesons, the QCD leading terms have much smaller overall coefficients, so that
the non-leading terms are expected to dominate at present energies $W<4\,GeV$, and the energy dependence is steeper:
$\sigma({\ov {M^{o}}}M^{o})\sim 1/W^{10}$.
This agrees with data on $\sigma({\ov K}_S K_s)$ and $\sigma(\pi^{o}\eta)$, \, while $\sigma(\pi^{o}\pi^{o})$
behaves "abnormally"  (may be due to contamination of data with the pure QED - background).

3)\,\, Predictions of the "standard handbag model"\, {\it disagree} with data either in energy and/or  angular dependence,
or in absolute values. However, it is not excluded that adding soft contributions from the 3-particle wave functions
in the "extended handbag model" (see Fig.4) can help to describe cross sections of {\it neutral mesons} at intermediate
energies $2.5\, GeV<W<4$. Unfortunately, such contributions are not yet calculated at present (and it well may be that
they are too small; besides, one has to remember that there are also power corrections due to the higher twist wave
function components in the diagrams in Fig.1\,).

\vspace*{3mm}

\begin{scriptsize}\vspace*{-3mm}
Table\,1\,: \,\, The value of \, "n" \, in $\sigma_{\rm tot} \sim (1/W)^{n}$ in
various reactions fitted in the $W$ and $|\cos\theta|$ ranges indicated.
\end{scriptsize}

\begin{center}\vspace*{0mm}
\hspace*{-.7 cm}
\begin{tabular}{l|| c|c|c|c||c|c} \hline \hline
Process & n - experiment & $W$ range (GeV) & $|\cos\theta|$ & Ref. & n - QCD & n - handbag \\
\hline\hline
$\pi^+\pi^-$ & $7.9 \pm 0.4_{\rm stat} \pm 1.5_{\rm syst}$ & $3.0 - 4.1$ & $<0.6$ & \cite{Nakaz} & $\simeq 6$  & $\simeq 10$ \\
\hline
$K^+K^-$  & $7.3 \pm 0.3_{\rm stat} \pm 1.5_{\rm syst}$ & $3.0 - 4.1$ & $<0.6$ & \cite{Nakaz} & $\simeq 6$ & $\simeq 10$ \\
\hline
$K^0_S K^0_S$  & $10.5 \pm 0.6_{\rm stat} \pm 0.5_{syst}$ & 2.4 - 4.0 & $<0.6$ & \cite{Chen} & $\simeq 10$ & $\simeq 10$ \\
\hline
$\eta \pi^0$ & $10.5 \pm 1.2_{\rm stat} \pm 0.5_{\rm syst} $ & 3.1 - 4.1 & $<0.8$ & \cite{Ue-eta} & $\simeq 10$ & $\simeq 10$ \\
\hline
$\pi^0\pi^0$ & $8.0 \pm 0.5_{\rm stat} \pm 0.4_{\rm syst}\,\, {\bf ?}$ & 3.1 - 4.1 & $<0.8$ & \cite{Ue-pi} & $\simeq 10$
& $\simeq 10$  \\
\hline\hline
\end{tabular}
\end{center}
\vspace{0mm}
\begin{minipage}[c]{0.4\textwidth}
{\hspace*{-0cm}\includegraphics
[trim=0mm 0mm 0mm 0mm, width=0.65\textwidth,clip=true]{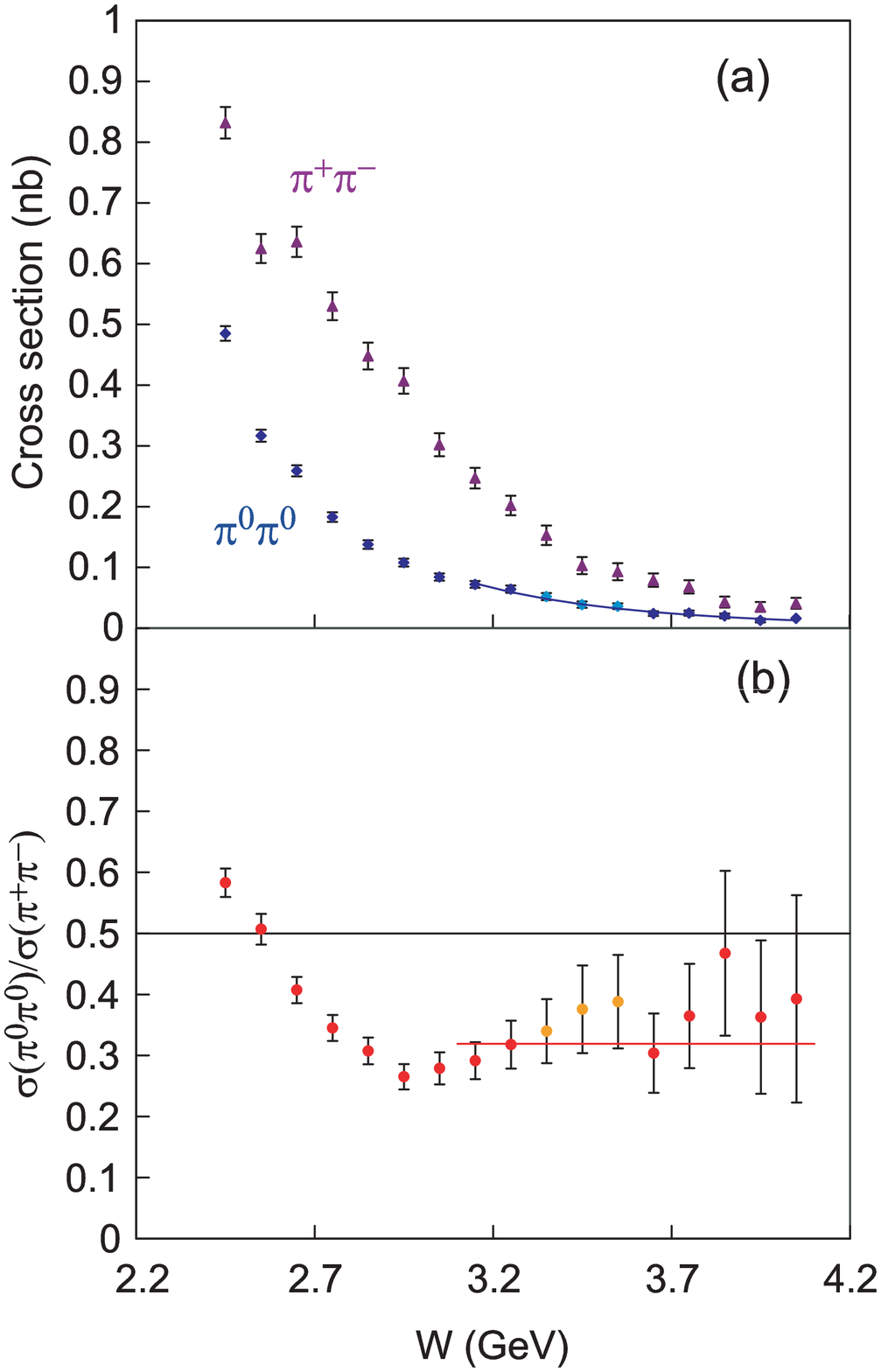}}

\begin{scriptsize}\hspace*{-2mm}
Fig. 7\quad  (a) Cross sections\,\, $\sigma_{o}(\gamma\gamma\to \pi^0\pi^0)$\\
\hspace*{-0.5cm} and $\sigma_{o}(\gamma\gamma\to \pi^+\pi^-)$ for $|\cos \theta^*|<0.6$\, \cite{Ue-pi, Nakaz}.\\
\hspace*{1cm}(b) Their\, ratio.
\end{scriptsize}
\end{minipage}
\begin{minipage}[c]{0.4\textwidth}{ \vspace*{-13mm}
\hspace*{-2.6cm}
\includegraphics[trim=0mm 0mm 0mm 0mm, width=1.1\textwidth,clip=true]{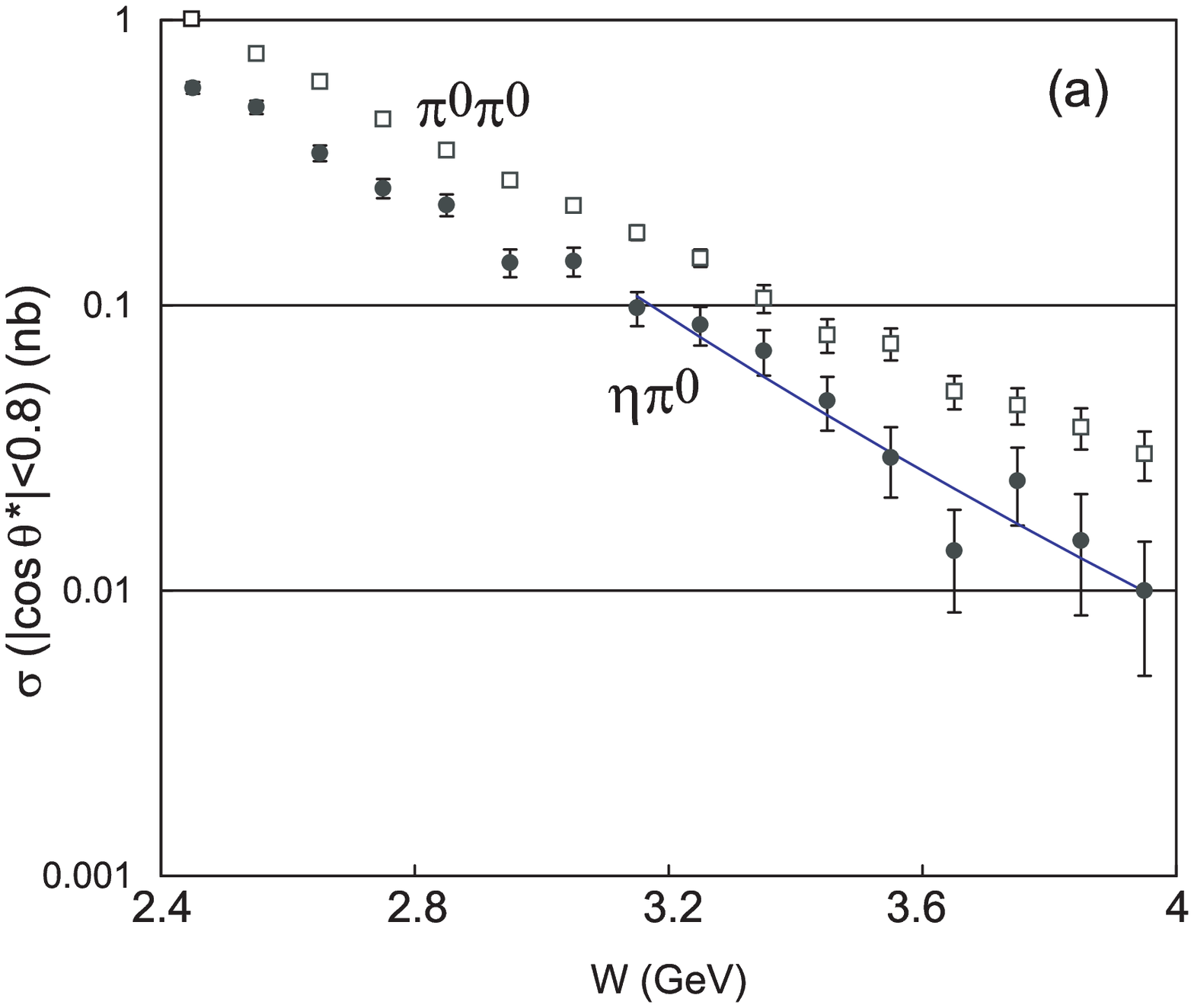}}
\end{minipage}
\begin{minipage}[c]{.4\textwidth}
\hspace*{-3.2cm}\includegraphics
[trim=0mm 0mm 0mm 0mm, width=0.8\textwidth,clip=true]{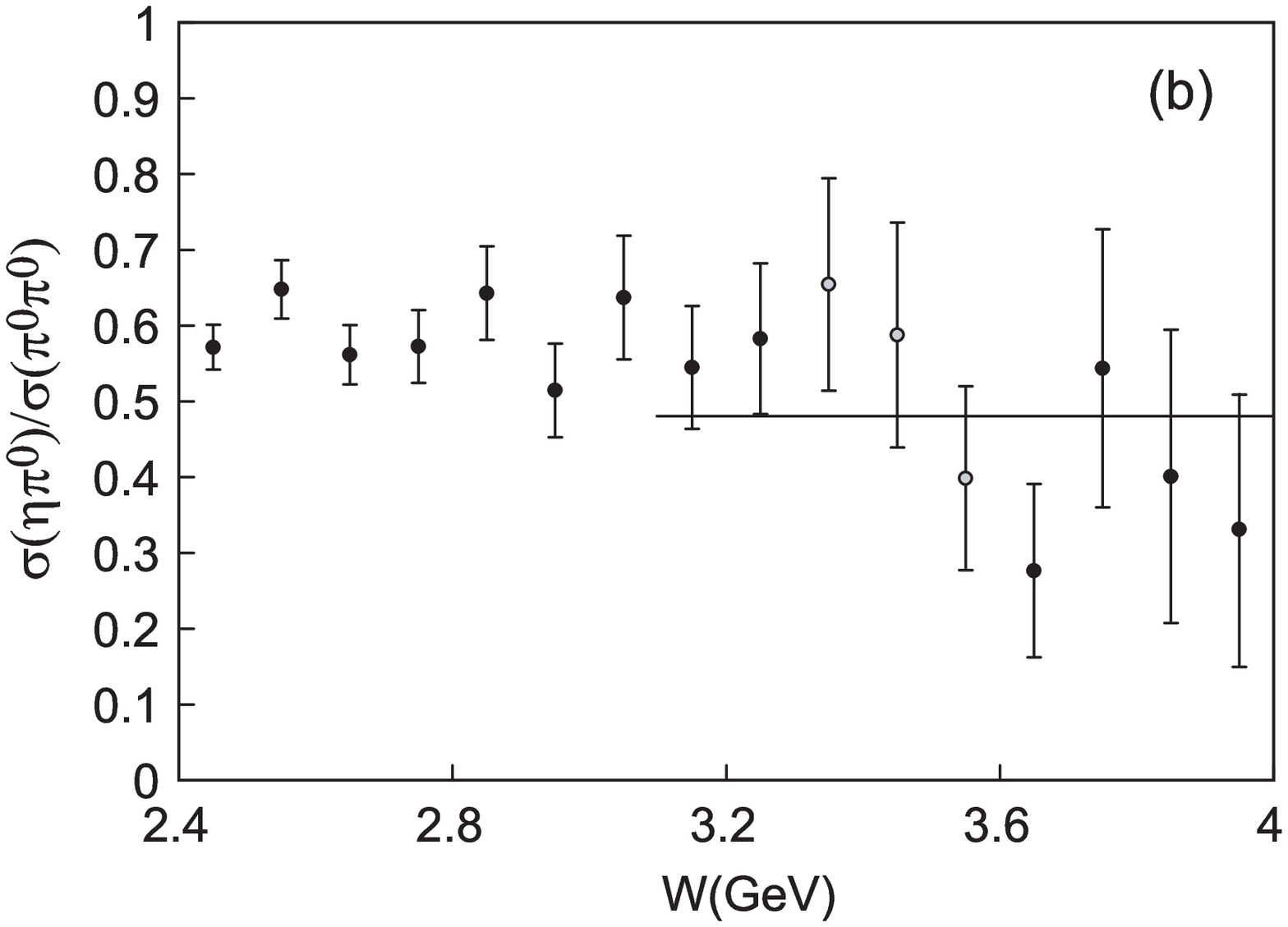}
\end{minipage}

\vspace*{-2cm}{
\begin{scriptsize}\hspace*{5cm}
Fig.8 \quad\,a)\,\,  $W$ - dependence of  cross sections $\gamma\gamma\ra \pi^{o}\pi^{o}$ \,
and $\gamma\gamma\ra \pi^{o}\eta\,\, \,(|\cos \theta^*|<0.8$). \\
\hspace*{6.5cm} The curve is the fit: \quad $ \sigma(\pi^o\eta)\sim W^{-n},\quad n=(10.5\pm 1.2 \pm 0.5)$\,
\cite{Ue-pi, Ue-eta}.\\
\hspace*{6.2cm} b)\,\,  $W$ - dependence of the cross section ratio  $\sigma(\eta\pi^0)
/\sigma(\pi^0\pi^0)\,\, (|\cos \theta^*|<0.8$).
\end{scriptsize}

\begin{center}
{\bf 4.} \quad $\Large \bm {\gamma^*\gamma\to P=\{\pi^o\,,\,\eta\,,\,\eta'\} }$ \, {\large\bf form factors} $\Large\bm{
F_{\gamma P}(Q^2)}$
\end{center}

As was first obtained in \cite{cz2} on the example of the pion form factor $F_{\pi}(Q^2)$ (see \cite{cz3} for
more details), the contributions from short and large distances factorize in $F_{\pi}(Q^2)$ at large $Q^2$, and the
logarithmic evolution of $Q^2 F_{\pi}(Q^2)$ is determined by renormalization factors of operators with the same
anomalous dimensions $\gamma_n$ as in the deep inelastic scattering. So, {\it the strict non-perturbative} QCD prediction
for $ F_{\pi}(Q^2)$ in the formal limit $Q^2\to \infty$ looks as \cite{cz2} ($b_o=11-2n_f/3,\, \mu_o\sim 1\,GeV$)\,:
\vspace*{-4mm}
\bq
F_{\pi}(Q^2)\to \frac{8\pi\alpha_s(Q^2))|f_{\pi}|^2}{Q^2}\Biggl (1+O\Bigl (\frac{\alpha_s(Q)}{\alpha_s(\mu_o)} \Bigr )^
{\frac{50}{9b_o}} \Biggr )= \frac{32\pi^2|f_{\pi}|^2}{b_oQ^2\ln Q^2}\Biggl (1+O\Bigl (\frac{\alpha_s(Q)}{\alpha_s(\mu_o)}
\Bigr )^{\frac{50}{9b_o}} \Biggr )\,.\nonumber
\eq
This corresponds to the pion wave function $\phi_\pi(x,\mu\to \infty)$ evolving to its universal asymptotic form
\vspace*{-4mm}
\beq
\phi_\pi(x,\mu\gg 1\,GeV)\to 6x(1-x)
\Biggl (1+O\Bigl (\frac{\alpha_s(\mu)}{\alpha_s(\mu_o)} \Bigr )^{\frac{50}{9b_o}} \Biggr )\,,\nonumber
\vspace*{-3mm}\eeq
{\it independently of its form $\phi_\pi(x,\mu_o\sim 1\,GeV)$ at low energy}.
As it is seen, the logarithmic evolution with increasing scale is very slow.

As for the form factor $F_{\gamma \pi}(Q^2)=F_{\gamma \pi}(Q^2=-q_1^2,\,q_2^2=0)$, the QCD prediction~
\footnote{\,
Really, unlike $F_{\pi}(Q^2)$, the leading asymptotic behavior of $F_{\gamma\pi}(Q^2)$ can be directly obtained from the
standard Wilson operator expansion of $J_{\mu}(z)J_{\nu}(0)\to \sum_n C_n(z) O_n(0)$ in (6) \cite{Ch3}, as in calculations of
the deep inelastic scattering. The only difference is that the forward matrix elements $\langle p|O_n|p\rangle$ are taken
in the deep inelastic scattering, while these are $\langle p|O_n|0\rangle$ in the case $\gamma^*\gamma\to \pi$.
}
for its asymptotic behavior in the formal limit $Q^2\to \infty$ looks as (see e.g. \cite{BL1})\,:
\beq
\int dz\, e^{iq_1 z}\langle \pi(p)|T\{J_{\mu}(z)J_{\nu}(0) \}|0\rangle=\Bigl (i\epsilon_{\mu\nu\lambda\sigma}
q_{1}^{\lambda}q_{2}^{\sigma}\Bigr )\,F_{\gamma \pi}(Q^2),
\eeq
\vspace*{-8mm}\hspace*{-2cm}
\beq
 Q^2 F_{\gamma \pi}(Q^2)=\frac{\sqrt 2\,f_\pi}{3}\int_0^1 dx\,\frac{\phi_\pi \Bigl (x,\mu\sim Q
 \Bigr )}{x}\Biggl (1+O \Bigl (\alpha_s(Q)\Bigr  ) \Biggr )=\sqrt 2\,f_\pi\,
 \Biggl (1+O\Bigl (\frac{\alpha_s(Q)}{\alpha_s(\mu_o)}\Bigr )^{\frac{50}{9b_o}} +O \Bigl (\alpha_s(Q)\Bigr  )\Biggr ).
\nonumber
\eeq

For the $\eta$ and $\eta'$ mesons, the form factors $F_{\gamma \eta}$ and $F_{\gamma \eta'}$ look similarly
to $F_{\gamma \pi}$. For instance, a simplified description of $|\eta\rangle,\,|\eta'\rangle$ states
in the quark flavor basis looks as follows~\cite{FKS}:
\beq
|\pi^o \rangle \to |({\ov u}u-{\ov d}d)/{\sqrt 2} \rangle,\quad
|n\rangle\to |({\ov u}u+{\ov d}d)/{\sqrt 2} \rangle,\quad  |s \rangle\to |{\ov s}s \rangle,\nonumber
\eeq
\vspace*{-4mm}
\beq
|\eta\rangle=\cos\phi \,|n\rangle-\sin\phi \, |s\rangle\,, \quad |\eta'\rangle=\sin\phi \,|n\rangle+\cos\phi \,
|s\rangle\,.
\eeq
\vspace*{-4mm}
\beq
f_{\pi}\simeq 132\, MeV,\quad f_{n}\simeq f_{\pi},\quad f_{s}\simeq 1.3\,f_{\pi},\quad \phi\simeq 38^{o}. \nonumber
\eeq
\beq
\hspace*{-2mm} F_{\gamma\pi}(Q^2)=
\frac{\sqrt 2 (e_u^2-e_d^2)\, f_{\pi}}{Q^2}\int_0^1 dx \,\frac{\phi_{\pi}(x,\mu\sim Q)}{x}\, I_o\, ;\,\, F_{\gamma n}(Q^2)=
\frac{\sqrt 2 (e_u^2+e_d^2)\, f_{\pi}}{Q^2}\int_0^1 dx \,\frac{\phi_{\pi}(x,\mu\sim Q)}{x}\, I_o\,,\nonumber
\eeq
\vspace*{-4mm}
\beq
F_{\gamma s}(Q^2)=
\frac{2 e_s^2 \, f_{s}}{Q^2}\int_0^1 dx \,\frac{\phi_{s}(x,\mu\sim Q)}{x}\, I_o\,; \quad I_o=\Biggl(1+O(\alpha_s)+
O(1/Q^2)\Biggr ),\,\, Q^2\gg 1\,GeV^2.
\eeq
\vspace*{-4mm}
\beq
F_{\gamma\eta}(Q^2)=\Biggl ( \cos \phi\, F_{\gamma n}(Q^2)-\sin \phi \,F_{\gamma s}(Q^2)\Biggr ),\,\,
F_{\gamma\eta'}(Q^2)= \Biggl (\sin \phi \,F_{\gamma n}(Q^2)+\cos \phi \, F_{\gamma s}(Q^2)\Biggr ).\nonumber
\eeq

Predictions for $F_{\gamma \pi}(Q^2)$  were given in a large number of theoretical papers, using many different
models for the leading twist pion wave function $\phi_{\pi}(x,\mu)$. The previous data for $F_{\gamma \pi}(Q^2)$\cite{cello,
Savin} covered the space-like region $0<Q^2<8\,GeV^2$. The recent data from BaBar \cite{Druz-pi, Druz-conf} extended
this one to $Q^2\lesssim 40\,GeV^2$, see Fig.9. It is seen that $Q^2 F_{\gamma P}(Q^2)$ exceeds its asymptotic value
$\sqrt 2\,f_\pi$ (the dashed line in Fig.9) at $Q^2\gtrsim 10\,GeV^2$. Because the loop and leading power corrections
are negative here, this shows that the leading twist pion wave function $\phi_\pi(x,\mu)$ is considerably wider than
$\phi_{\rm asy}(x)$, while most theoretical models predicted $\phi_\pi(x,\mu)\simeq \phi_{\rm asy}(x)$. The red curve in
Fig.9 shows that, with power corrections of reasonable size, the wide leading twist pion wave function $\phi_{\rm cz}(x,\mu)$
obtained in \cite{cz} using the standard QCD sum rules, see Fig.2, is not in contradiction with data.
\footnote{\,
A number of papers with predictions for $F_{\gamma \pi}(Q^2)$ has been published, based on the model pion wave function
$\phi_{\pi}^{\rm BMS}(x)$ (BMS=Bakulev-Mikhailov-Stefanis) obtained from the "improved QCD sum rules" with non-local
condensates (see the last paper \cite{MS} and references therein). This approach has been criticized in \cite{Ch2}, as it is
based on arbitrary strong dynamical assumptions which, as was shown in \cite{Ch2}, don't pass the direct QCD check. Moreover,
within this approach, one has to introduce {\it a number of arbitrary model functions} for various non-local vacuum condensates
(see e.g. \cite{BP}) and, in general, the results for $\phi_{\pi}(x)$ depend heavily on the model forms chosen for these
functions (compare e.g. the results for $\phi_{\pi}(x)$ from \cite{Radyu} and \cite{BP}). Finally, the model pion wave function
$\phi_{\pi}^{\rm BMS}(x,\mu)$ obtained within this approach predicted the value of $F_{\gamma \pi}(Q^2)$ only slightly above
those for the asymptotic wave function $\phi_{\pi} (x)=\phi_{\rm asy}(x)$, see \cite{MS}, and well below the recent BaBar
data \cite{Druz-pi}\cite{Druz-conf}.
Besides, it is claimed in \cite{MS} that the data \cite{cello}\cite{Savin}\cite{Druz-pi}\cite{Druz-conf} are incompatible
with $\phi_{\pi}(x,\mu)=\phi_{cz}(x,\mu)$ (and are even in contradiction with the QCD factorization for any pion wave function
with the end point behavior $\sim x(1-x)$ at $x\to 0,\,1$\,). As it is seen from Fig.9 (red curve), this is not so.
\vspace*{2mm}
}

After the new BaBar data on $F_{\gamma \pi}(Q^2)$ \cite{Druz-pi} appeared, it was proposed in \cite{Rad} that the large
value of $Q^2 F_{\gamma \pi}(Q^2)$ is due to the flat pion wave function, $\phi_{\pi}(x,\mu\sim 1\,GeV)\simeq 1$, as $Q^2
F_{\gamma \pi}(Q^2)$ grows $\sim \ln (Q^2/M^2)$ in this case
and, taking $M \simeq m_{\rho}$ by hand, such a behavior fits well then these BaBar~ data.
But in this case, because the wave functions of $|\pi\rangle,\, |n\rangle$ and $|s\rangle$ are qualitatively similar, see
(7-8), the form factors $q^2F_{\gamma\eta}(q^2)$ and  $q^2F_{\gamma\eta'}(q^2)$ will also grow the same way, $\sim \ln
(q^2/m^2_{\rho})$ at $q^2\gg 1\, GeV^2$. These two form factors have been measured recently by the BaBar Collaboration
\cite{Druz-eta} at $q^2=112\,GeV^2$. It is seen from Table 2 and Fig.10 that with $\phi_{n}(x)\sim \phi_{s}(x) \simeq 1$
these two form factors will be too large. Besides, such flat wave functions will contradict the data on  $\sigma(e^+e^-\to
VP)$, see Fig.11\,.

\begin{footnotesize}
\vspace*{7mm}
\hspace*{-0.5cm}
{Table 2. \, The  values of form factors $|q^2F_{\gamma P}(q^2)|$ (in $GeV$) at $q^2=112\,GeV^2$ for various meson
wave functions}
\end{footnotesize}~

\vspace*{8mm}
\hspace*{0.5cm}{
\renewcommand{\arraystretch}{2.0}
\begin{footnotesize}
\begin{tabular}{l|c|c|c|c} \hline \hline

Wave functions & $|q^2 F_{\gamma^{*}\pi}(q^2)|$ & $|q^2 F_{\gamma^{*}\eta}(q^2)|$ & $|q^2 F_{\gamma^{*}\eta^\prime}(q^2)|$  &
Ref.\\ \hline \hline

$\phi_{n}(x)\simeq \phi_{s}(x)\simeq \phi_{asy}(x)=6 x(1-x)$ &  $0.14$  & $0.13$ & $0.21$ &

\\ \hline

$\phi_{n}(x)\simeq \phi_{s}(x)\simeq \phi_{cz}(x)$  & $0.22$ & $0.21$ & $0.33$ &
\\ \hline

$\phi_{\bf n}(x)\simeq\phi_{\bf cz}(x);\quad \phi_{\bf s}(x)\simeq\phi_{\bf asy}(x)$ & $\bm {0.22}$ & $\bm{0.24}$
&$\bm{ 0.29 }$&

\\ \hline \hline

$\phi_{n}(x)\simeq \phi_{s}(x)\simeq 1$ & $0.32$ & $0.31$ & $0.49$ &

\\ \hline \hline

experiment & --- & $\mathbf{0.23\pm 0.03}$ & $\mathbf{0.25\pm 0.02}$ & \cite{Druz-eta}\\
\hline\hline
\end{tabular}
\end{footnotesize}
}
\newpage
\begin{minipage}[c]{.7\textwidth}
\vspace{3mm}\hspace*{-20mm}
\includegraphics[width=0.65\textwidth]{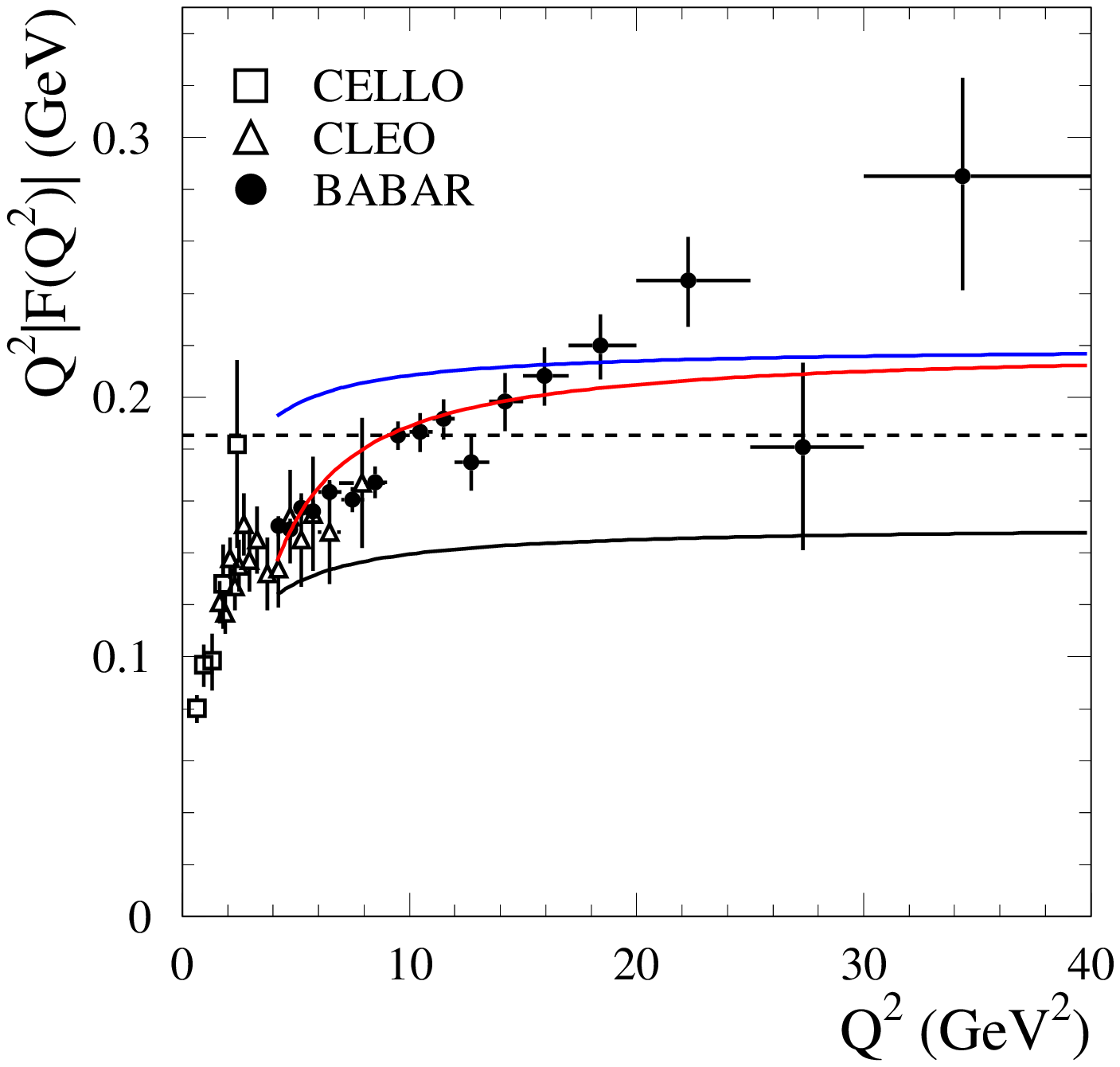}
\end{minipage}~
\begin{minipage}[c]{0.45\textwidth}\vspace{-1.5cm}
\begin{scriptsize}\vspace{+2cm}
{\hspace*{-5cm} Fig.9\quad The form factor $\Phi\equiv Q^2F_{\gamma\pi}(Q^2)$}\\

\hspace*{-2cm}Theory\,: \vspace*{2mm}\\
\hspace*{-6.4cm} \,a)\,\, the logarithmic loop corrections are calculated (in part) at the NNLO \,\,
\cite{Kivel}\cite{Melic}\,;\\

\hspace*{-6.4cm} b)\,\, only the part of the total power correction $\sim 1/Q^2$ is calculated
at present in \cite{Khodja}\,:\\

\hspace*{-3cm} $\delta \Phi_4\simeq -{\sqrt 2}f_{\pi}(0.6\, GeV^2)/Q^2$\,,\\

\hspace*{-6.4cm} originating from the 2 -  and 3 - particle {\bf asymptotic} pion wave functions of twist 4 \cite{Br-Fil}.\\
\hspace*{-6.4cm} It well may be that it is not even the main part of the total $\sim 1/Q^2$ correction, because: \\
\hspace*{-6.4cm} i) the deviation of these 2 - and 3 - particle twist 4 pion wave functions from their asymptotic \\
\hspace*{-6.4cm} forms (i.e. the admixture of higher "partial waves") was estimated in \cite{Br-Fil} only for the first \\
\hspace*{-6.4cm} non-leading "partial wave", while second (and higher) "partial waves" neglected in \cite{Br-Fil} can be \\
\hspace*{-6.4cm} important, \,  ii) there are also contributions $\sim 1/Q^2$ from the 4 -particle wave functions of \\
\hspace*{-6.4cm} twist 4,\, iii) moreover, the twist expansion breaks down at this level, so that the higher \\
\hspace*{-6.4cm} twist  $\geq 6$ terms also give  contributions $\sim 1/Q^2$\,;\\

\vspace{2mm}
\hspace*{-6.4cm} c)\,\, the power correction $\sim 1/Q^4$ is unknown.\\

{\hspace*{-5.5cm} Black line:\,  $\phi_{\pi}(x)=\phi_{\rm asy}(x),\quad \Phi\simeq {\sqrt 2}f_{\pi}\Bigl
[\, 0.77-(0.6\, GeV^2/Q^2)\,\Bigr ]$ \, \cite{Kivel}\cite{Melic}\cite{Khodja}}\\
{\hspace*{-5.5cm}  Blue line\, :\,  $\phi_{\pi}(x)=\phi_{\rm cz}(x),\quad\quad \Phi\simeq {\sqrt 2}
f_{\pi}\Bigl [\, 1.18-(0.6\, GeV^2/Q^2) \,\Bigr ]$ \cite{Melic}\cite{Khodja}}\\
{\hspace*{-5.5cm} {\bf Red line} (the example with additional mild power corrections)\,: }\\
{\hspace*{-6.0cm}  $\bm{\phi_{\pi}(x)=\phi_{\rm cz}(x),\quad \Phi\simeq {\sqrt 2} f_{\pi}\Bigl [\,
1.18-(1.5\, GeV^2/Q^2)-(1.2\, GeV^2/Q^2)^2 \,\Bigr ]}$\,. }\\
\end{scriptsize}
\vspace*{2mm}{\hspace*{-2.5cm}Experiment\,: \cite{cello}\cite{Savin}\cite{Druz-pi}\cite{Druz-conf}}
\end{minipage}

\vspace*{-1mm}
\begin{minipage}[c]{.5\textwidth}
\vspace{3mm}
\includegraphics
[trim=0mm 0mm 0mm 0mm, width=0.6\textwidth,clip=true]{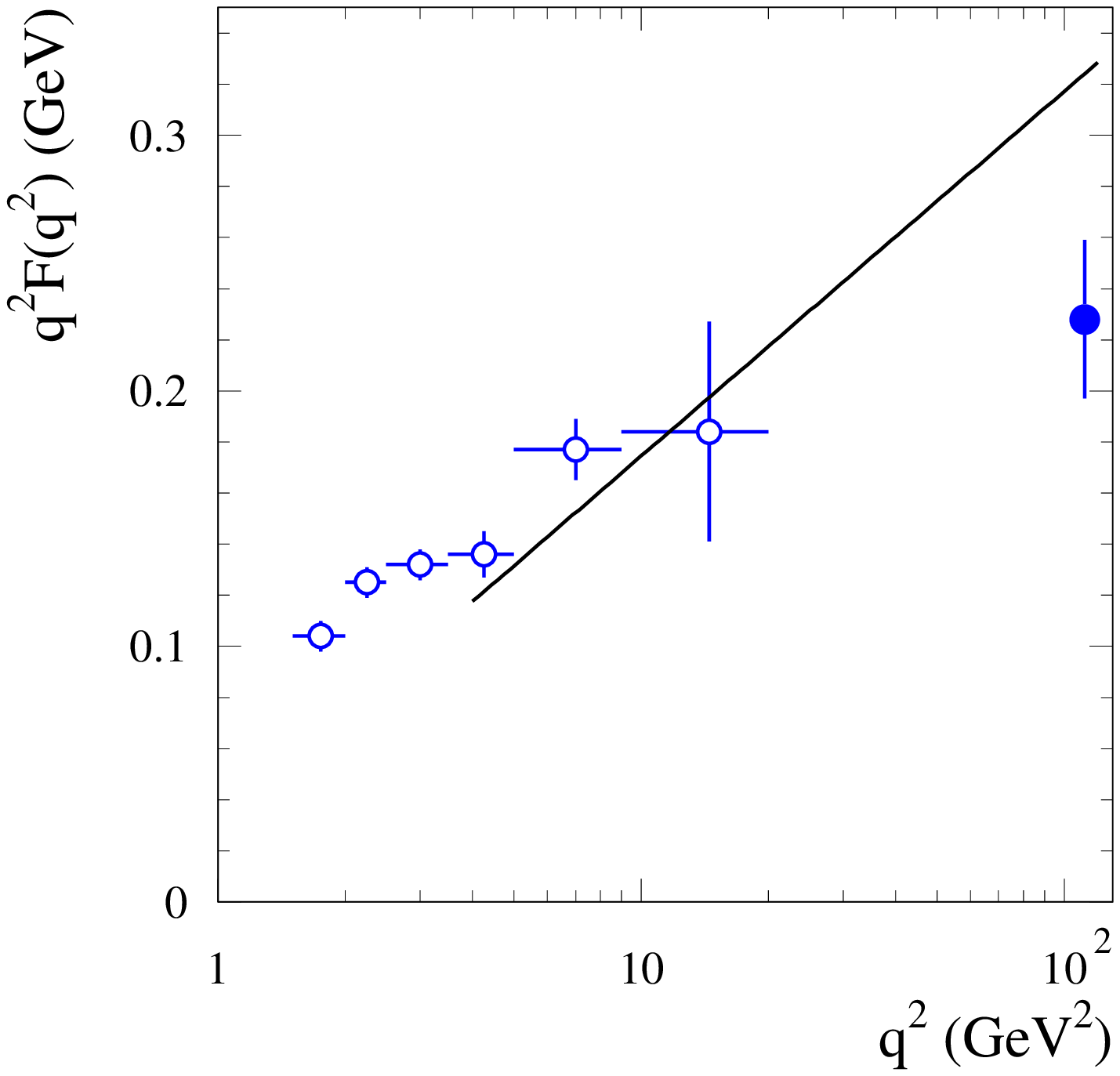}
\end{minipage}~
\begin{minipage}[c]{.5\textwidth}
\vspace{6mm}
\includegraphics
[trim=0mm 0mm 0mm 0mm, width=0.6\textwidth,clip=true]{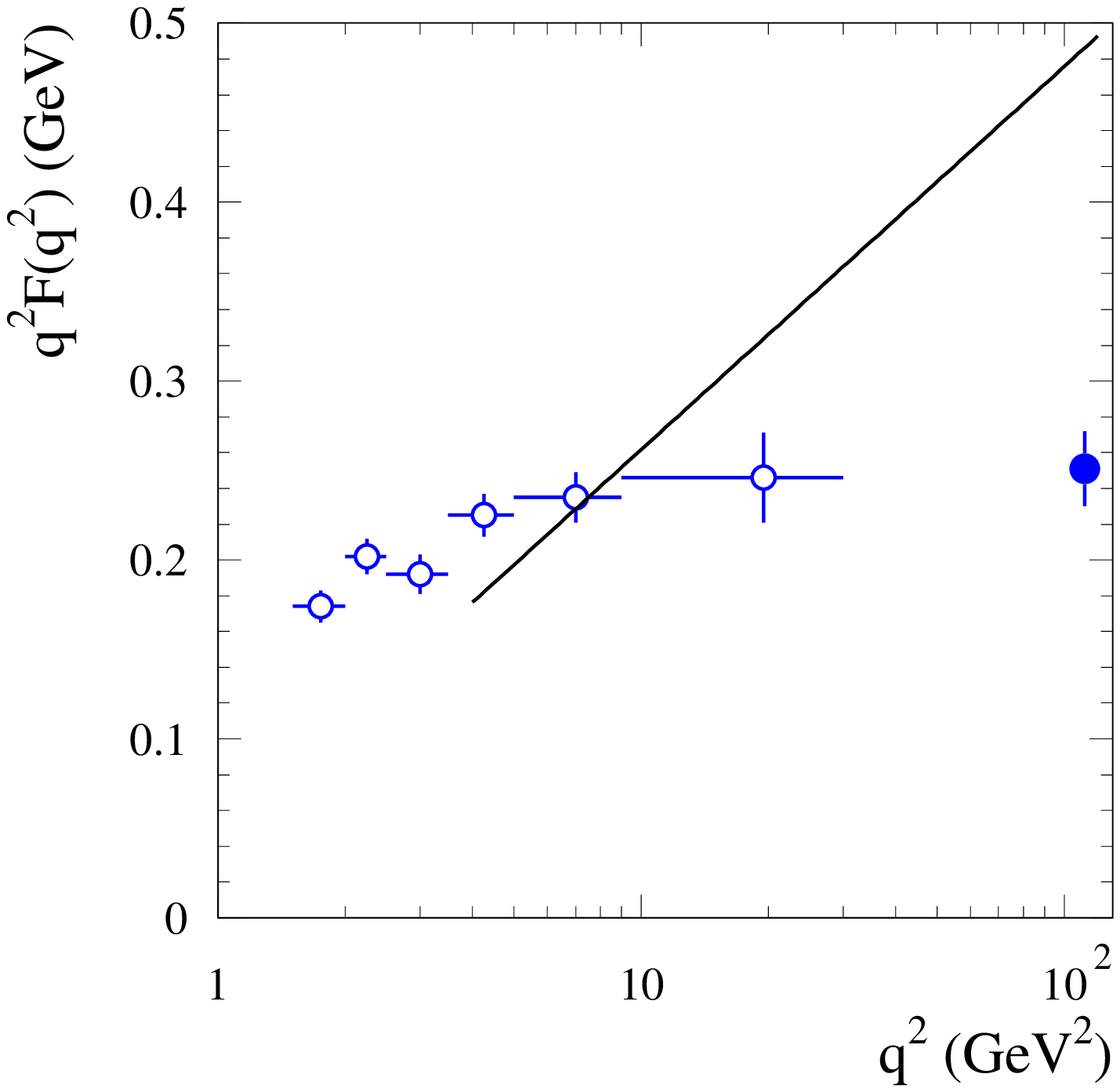}
\end{minipage}~
\vspace{-2mm}
\begin{footnotesize}
Fig.10\,\,\,\quad\quad Full\, points:\, \, $|q^2F_{\gamma\eta}(q^2)|$ (left) and\, $|q^2F_{\gamma\eta'}(q^2)|$ (right)
transition form factors at $q^2=112 \, GeV^2$\, \cite{Druz-eta}:\\
${\hspace{3cm}} |q^2 F_{\gamma\eta}(q^2)|=(0.229\pm 0.030\pm 0.008)\,GeV,\,
\quad  |q^2 F_{\gamma\eta^{\prime}}(q^2)|=(0.251\pm 0.019\pm 0.008)\,GeV$.\\
\vspace*{-6mm}{
\begin{center}
White\, points:\,\, previous CLEO data at \,\,$2\,GeV^2<(Q^2=-q^2)<20\,GeV^2$\, \cite{Savin}.\,\,\\
\vspace{1mm}
Black lines:\, the form factors $|q^2 F_{\gamma\eta}(q^2)|,\,\,|q^2 F_{\gamma\eta'}(q^2)|$
for the flat pseudoscalar wave function $\phi_{P}(x)\simeq 1$.
\end{center}
}
\end{footnotesize}

\vspace{2mm}
\begin{center}
{\bf 5.} \quad {\large \bf Conclusions on the form  factors} $\bm{F_{\gamma P}(Q^2),\, P=\{\pi^o,\,\eta,\,\eta' \}}$
\end{center}
\begin{center}{\vspace{-3mm}}
   {\large\bf and the leading twist wave functions\, $\bm{\phi_P(x)}$ of pseudoscalar mesons}
\end{center}
\vspace{-1mm}

The {\it flat leading twist pseudoscalar wave function}\,\, $\phi_{P}(x)\simeq 1$ :\\
a)\, predicts the form factors $F_{\gamma \eta}(q^2)$ and $F_{\gamma \eta'}(q^2)$ at $q^2=112\, GeV^2$
considerably larger than the BaBar results;\\
b)\, predicts the parametrical behavior of cross sections $\sigma(e^{+}e^{-}\ra VP)$ at large
$s$ as\,:\, $\sigma(e^{+}e^{-}\to VP)\sim 1/s^2$\,, in contradiction with the data
$\sigma(e^{+}e^{-}\to VP)\sim 1/s^4$ in the interval $\sim 8\, GeV^2<s<112\, GeV^2$.

\vspace{2mm}
The {\it asymptotic leading twist pseudoscalar wave function}\,\, $\phi_P(x)\simeq 6x(1-x)$ :\\
a)\, predicts the form factors $F_{\gamma \pi^o}(Q^2),\, F_{\gamma \eta}(Q^2)$ considerably smaller
than data;\\
b)\, predicts branchings of charmonium decays: ${^3}P_{0},\, {^3}P_{2}\to \pi^{+}\pi^{-},\, K^{+}K^{-}$,
the pion and kaon electromagnetic form factors $F_{\pi,K}(q^2)$ at $q^2=10-15\, GeV^2$ much smaller than data, etc.
\vspace{1mm}

The {\it CZ leading twist pion wave function}\,\,$\phi_{\pi}^{cz}(x,\mu\sim 1\,GeV)=30x(1-x)(2x-1)^2$\,:\\
\,leads to predictions which, it seems, are not in contradiction with all data available.\\

\vspace*{-1cm}
\begin{minipage}[c]{.55\textwidth}{\hspace{-15mm}}
\includegraphics[width=1.1\textwidth]{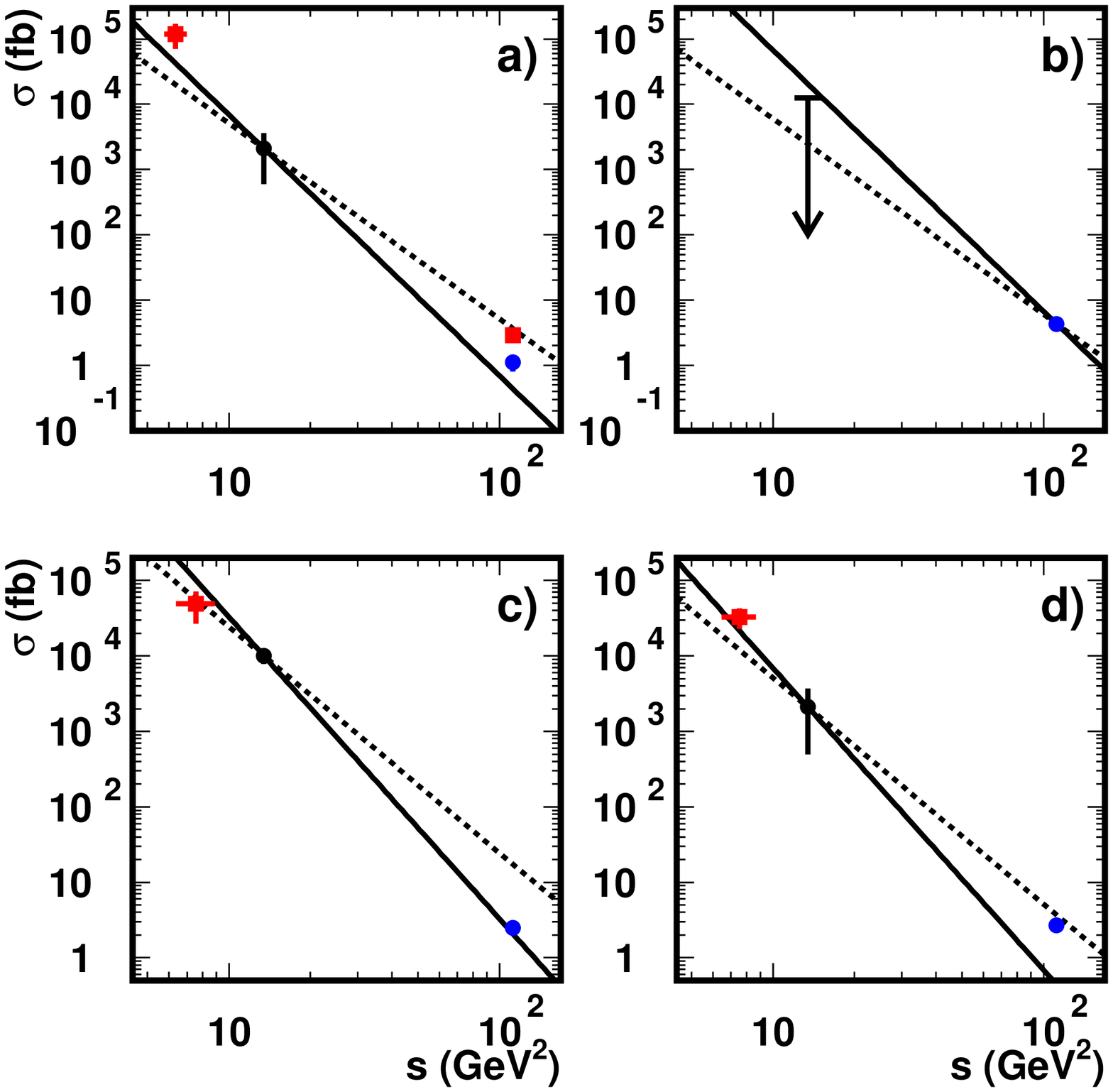}
\end{minipage}~
\begin{minipage}[c]{.65\textwidth}\vspace{-2cm}
\begin{footnotesize}\vspace*{4.5cm}{
{\hspace*{0cm} Fig.11\quad Solid lines correspond to $1/s^4$ dependence}\\
{\hspace*{0.5cm} and  dashed ones represent $1/s^3$}.\\

\vspace*{3mm}

{\hspace*{1cm} {\bf a)} $\sigma(e^+e^- \to {\phi \eta})$\quad {\bf b)} $\sigma(e^+e^- \to {\phi \eta'})$}\\

{\hspace*{12mm}{\bf c)} $\sigma(e^+e^- \to {\rho \eta})$\quad {\bf d)} $\sigma(e^+e^- \to {\rho \eta'})$ }\\

The measured cross sections: \\
at $\sqrt{s}\simeq 2.5,\,2.75\, GeV$ by BaBar \cite{BB-1},\\
at $\sqrt{s} = 3.67 $ GeV by CLEO \cite{CL},\\
at $\sqrt{s} = 10.58$ GeV by BaBar \cite{BB-2} and Belle \cite{VP}\\
for  various processes.\\
BaBar measurements are represented by squares .\\
}
\end{footnotesize}

\begin{scriptsize}
\vspace*{2mm}{
{\hspace*{0.5cm} QCD\, predictions\,\, :\,}\\

$\sigma(e^+e^- \to VP)\sim 1/s^4$\, \cite{cz1}\, (\, see (1), $|\lambda_V|=1$ in this case\,),\\
up to a possible additional logarithmically growing factor \cite{cz-r},\\
for the pseudoscalar  wave function $\phi_{P}(x)$ with the suppressed \\
end point behavior, like $\sim x(1-x)$ at $x\to 0,1$.\\

$\sigma(e^+e^- \to VP)\sim 1/s^2$\,\, for the flat pseudoscalar \\
wave function $\phi_{P}(x)\sim 1$.
}
\end{scriptsize}
\end{minipage}

\vspace{3mm}
\begin{scriptsize}
The form factors $\gamma^*\to VP$ are highly sensitive to the end point behavior of the leading twist pseudoscalar wave
function $\phi_P(x)$, as they contain the factor $I\sim \int_{\delta}^1 dx\,\phi_P(x)/x^2,\,\,\delta=O(\mu^2_o/Q^2)$\,\,
\cite{cz-r}. So, $I\sim \ln(Q^2/\mu^2_o)$ for $\phi_P(x)\sim x(1-x)$, while it will be parametrically larger
at $\phi_P(x)\sim 1$\, : $I\sim (Q^2/\mu^2_o)$.
The data are in a reasonable agreement (with a logarithmic accuracy) with the $\sigma\sim 1/s^4$ dependence corresponding
the end point behavior $\phi_P(x)\sim x(1-x)$, and are in contradiction with the behavior $\sigma\sim 1/s^2$, corresponding
to $\phi_P(x)\sim 1$ at $x\to 0,\, 1$.
\end{scriptsize}

\vspace*{3mm}\hspace*{3cm}
{\bf Note added}
\vspace*{2mm}

After this talk has been given, there appeared the paper \cite{DK} with updated "predictions" of the handbag model for the
$\gamma\gamma\to {\ov M}M$ cross sections. In comparison with the previous paper \cite{DKV}, the main new element in \cite{DK}
is that the sizeable soft non-valence form factor $R^{\rm nv}_{\ov M M}(s)$ is used now, in addition to the soft valence
one, $R^{\rm v}_{\ov M M}(s)$.
\footnote{\,
As for the "prediction" of the angular behavior $s\,d\sigma /{d \cos \theta} \sim |R_{{\ov M}M}|^2/{\sin^4 \theta}$ in \cite
{DKV}, see footnote 1 and \cite{Ch1}.
}
Both functions, $R^{\rm nv}_{{\ov M}M}(s)$ and $R^{\rm v}_{{\ov M}M}(s)$, are parameterized then in arbitrary forms, with
a large number of free parameters which are fitted to the data.
\footnote{\, The form factors $R^{\rm u}_{2\pi}(s)$ and $R^{\rm s}_{2\pi}(s)$ used in \cite{DK} are: $R^u_{2\pi}(s)=R^{\rm v}_
{2\pi}(s)+R^{\rm nv}_{2\pi}(s)$,\,\, $R^s_{2\pi}(s)=R^{\rm nv}_{2\pi}(s)$.
}
It seems, the authors consider that nearly nothing is known
about QCD, so that it is possible to proceed in such an arbitrary way.

As for the standard soft valence contributions to the cross sections and the soft valence form factors $R^{\rm v}_{{\ov M}M}
(s)$, these were estimated in \cite{Ch1} with a help of the QCD light cone sum rules (see section 2 above) and were found
much smaller (and with the behaviour $\sim 1/s^2$) than the values fitted in \cite{DKV} (and in \cite{DK}). In addition,
we would like to comment here in short on the non-valence contributions.

Two types of non-valence contributions are presented in Fig.12\,.

\vspace*{2mm}
\begin{minipage}[c]{.5\textwidth}{\hspace*{-5mm}}
\includegraphics[width=0.9\textwidth]{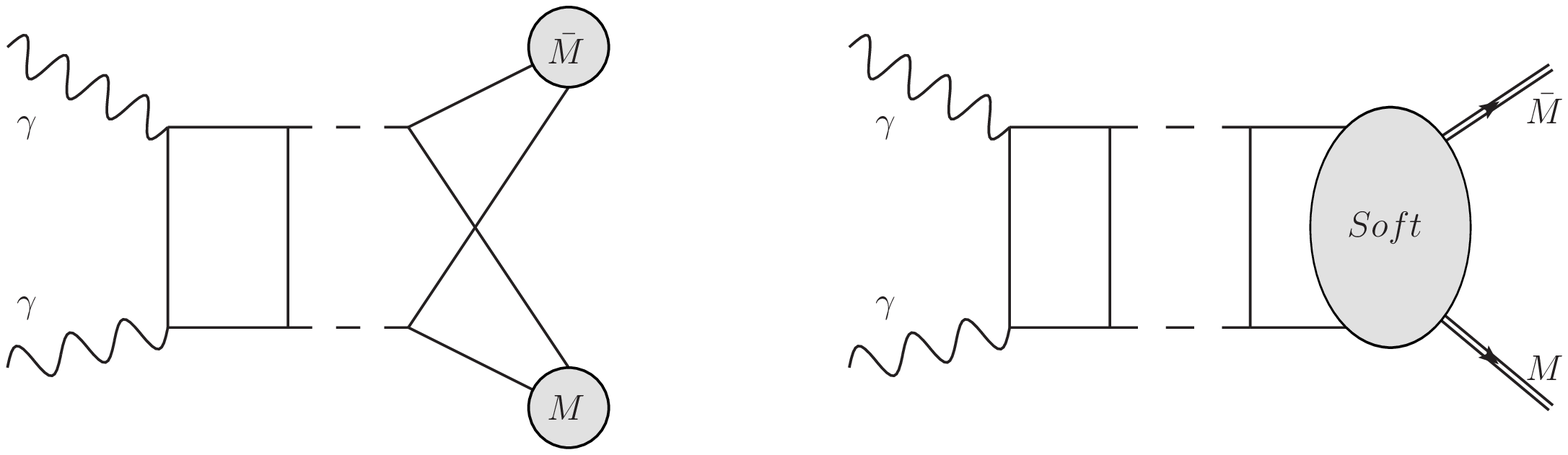}
\end{minipage}
\begin{minipage}[c]{.5\textwidth}
\begin{scriptsize}
\hspace*{-8mm} Fig.12a \quad The leading power non-valence one-loop correction. \\
\hspace*{-8mm} Fig.12b \quad The contribution to the soft handbag form factor $R^{\rm nv}_{{\ov M}M}(s)$.\\
\hspace*{-8mm} The solid and dashed lines represent quarks and gluons.
\end{scriptsize}
\end{minipage}

\vspace*{2mm}
It is worth noting that both non-valence contributions in Fig.12 are $SU(3)$-flavor singlets in the $SU(3)$-symmetry
limit. So, they contribute equally to the amplitudes $\pi^+\pi^-,\,\pi^0\pi^0,\,K^+K^-,\, \ov{K^0}K^0$ and 
 $\eta_8\eta_8$, and don't contribute to $\eta_8\pi^0$.

The diagrams in Fig.12a constitute a small subset of all one-loop corrections to the leading power contributions from
the Born diagrams like those shown in Fig.1\,. We only note here that if these leading power one-loop non-valence corrections
to the Born contributions were really significant, this will contradict then the data \cite{Chen} on $K_SK_S$, see Fig.6\,.~
\footnote{\,
In particular, this non-valence one-loop correction was calculated, among all others, in \cite{DN}. Its contribution into
$\sigma(\gamma\gamma\to K^+K^-)$ cross section (integrated over $|\cos\theta|<0.6$, and with $\phi_K(x)=\phi_{\rm asy}(x)$ )
is \cite{Dupl}\,: $\delta\sigma^{\rm nv}/\sigma \simeq (-\alpha_s(s)/3\pi)\simeq -3\%$.\, I.e., its (average) contribution
into the amplitude is: $\delta{\ov A}^{\rm nv}/{\ov A}(K^+K^-)\simeq -1.5\%$. The amplitude $|{\ov A}(K_S K_S)|\simeq 0.15\,
|{\ov A}(K^+K^-)|$, see BL and BC lines in Fig.6. So, the rough estimate of this non-valence one loop correction to the
${\ov A}(K_S K_S)$ amplitude in (2) is\,: $|\delta{\ov A}^{\rm nv}/{\ov A}(K_S K_S)|\simeq 10\%$.
}

As for the soft non-valence handbag form factor $R^{\rm nv}_{{\ov M}M}(s)$, it seems sufficient to say that it originates
first from  the Fig.12b two-loop (non-logarithmic) correction, so that\,: $R^{\rm nv}_{{\ov M}M}(s)/R^{\rm v}_{{\ov M}M}(s)
=O\Bigl ((\alpha_s(s)/\pi)^2\Bigr )=O(10^{-2})$\,. Clearly, so small non-valence contribution will not help.
\vspace{3mm}

I am grateful to V.P. Druzhinin for explaining me details of various experimental results and G.~Duplancic for providing me
with the additional details of calculations performed in \cite{DN}.
\vspace{2mm}

This work is supported in part by the RFBR grant 07-02-00361-a.\\

\end{document}